\documentstyle[twocolumn,epsfig,prl,aps,amsmath,amssymb,float]{revtex}
\begin{document}
\newfloat{figure}{ht}{aux}
\draft
\twocolumn[\hsize\textwidth\columnwidth\hsize\csname
@twocolumnfalse\endcsname
\title{Magnetization profiles and NMR spectra of doped Haldane chains at finite temperatures}
\author{F.~Alet and E.~S.~S\o rensen}
\address{Laboratoire de Physique Quantique \& 
UMR CNRS 5626, Universit\'e Paul Sabatier, 31062 Toulouse, France}  
\date{\today}
\maketitle
\begin{abstract}
Open segments of $S=1$ antiferromagnetic spin chains are studied at finite temperatures
and fields using continuous time Quantum Monte
Carlo techniques. By calculating the resulting magnetization profiles
for a large range of chain lengths with fixed field and temperature we
reconstruct the experimentally measured NMR spectrum of impurity
doped Y$_2$BaNi$_{1-x}$Mg$_x$O$_5$. For temperatures above the gap the
calculated NMR spectra are in excellent agreement with the experimental
results, confirming the existence of $S=1/2$ excitations at the end of
open $S=1$ chain segments. At temperatures
below the gap, neglecting inter chain couplings, we still find well defined
peaks in the calculated NMR spectra corresponding to the $S=1/2$ chain
end excitations. At low temperatures, inter chain couplings could become
important, resulting in a more complicated phase.
\end{abstract}
\vskip2pc]
\section{Introduction}
The physics of low-dimensional spin systems is extremely rich and shows
many surprising effects. Using a valence bond picture it has been
suggested that open segments of $S=1$ spin chains 
should have $S=1/2$
excitations localized at the ends of the chain~\cite{Kennedy,hagiwara1},
that are experimentally observable.
In a recent experiment by Tedoldi et al~\cite{Tedoldi}
the presence of these $S=1/2$ chain end spins was beautifully  demonstrated 
by measuring the NMR profile of Mg doped Y$_2$BaNiO$_5$, a model $S=1$
spin chain system, showing well
defined satellite peaks indicative of the chain end excitations. These satellite
peaks were, perhaps surprisingly, very well defined at temperatures
close to and well above the gap, $\Delta\simeq 100K$. At temperatures below the gap,
where one would expect the signal from the chain end spins to be
most clearly observable, the
satellite peaks were smeared into a single broad line. The inter chain
coupling $J^\perp$ in Y$_2$BaNiO$_5$ is very small 
$(|J^\perp/J|\le 5\times 10^{-4}, J\simeq 285K)$~\cite{Tedoldi}
and it therefore remains a question why no well defined satellite peaks
were observed below the gap.
Bulk susceptibility and magnetization
measurements~\cite{Payen_Chi} on Zn doped Y$_2$BaNiO$_5$ have also been 
interpreted in terms of $S=1/2$ chain end excitations, but in this case
at temperatures
all the way down to 4-10K, well below the gap. 
Very interestingly, a sub-Curie power law divergence was observed
at temperatures below 4K hinting at the presence of a gapless phase.
In light of these two experiments it is then natural to ask, from
a theoretical point of view, if chain end excitations are observable at
finite fields and temperatures well above the gap and when a simple
one-dimensional model will fail to describe the experiments.
In the present paper we systematically address these issues 
by calculating the on-site magnetization, $S^z_i$, at finite temperatures and
fields using continuous time quantum Monte Carlo techniques and
reconstructing the associated NMR spectra for a given concentration of
impurities.

It is by now well established that integer spin chains 
display a gap above a spin liquid ground-state~\cite{Haldane} as 
opposed to half-integer spin chains which are gapless. 
The gap in integer spin chain systems has been observed experimentally
in numerous compounds~\cite{GappedCompounds} as well as in many
numerical studies~\cite{numgap}. For $S=1$ chains the gap is known
to be $\Delta\sim 0.4105J$~\cite{white1}.
For integer spin chains, described by a simple antiferromagnetic
Heisenberg model, it is possible to understand the low temperature
physics using the intuitive valence bond picture. From this point
of view the ground-state is close to the valence bond solid (VBS)
state. Each integer spin $S$ is decomposed into  $2S$ spin $S=1/2$ in
the symmetric state and
each of the $S=1/2$ on neighboring sites are paired into singlets.
It is possible to determine Hamiltonians where this VBS
state is the exact ground-state. For $S=1$ spin chains the corresponding
Hamiltonian is rather simple and apart from the usual Heisenberg
term includes a biquadratic term $-\beta({\bf S}_i\cdot {\bf S}_{i+1})^2$.
For $\beta=-1/3$, the AKLT point, the VBS state is the exact
ground-state as was shown by Affleck-Kennedy-Lieb and Tasaki~\cite{AKLT} (AKLT).
Under periodic boundary conditions the VBS state is always a singlet.
However, as was noted by Kennedy~\cite{Kennedy} for open $S=1$ chains, the
ground-state is four-fold degenerate in the thermodynamic limit. In the
VBS picture this is easy to understand since the $S=1/2$ spins at the
end of the chain are not paired into a singlet state, leading to a four-fold
degeneracy for a $S=1$ system. In general a spin-$S$ chain should then
have spin-$S/2$ chain-end excitations.
Many real compounds are well described by
the antiferromagnetic Heisenberg chain:
\begin{equation}
H=J\sum_i{\bf S}_i\cdot{\bf S}_{i+1}.
\end{equation}
Following the above discussion the ground-state should then be close
to the VBS state. For a $S=1$ system, the low-energy states of an open
segment is then a singlet and a triplet separated by a gap decaying 
exponentially with L, the length of the segment. For L odd the interaction 
between the two chain end excitations is ferromagnetic and the triplet is the
lowest lying state, for L even the interaction is antiferromagnetic
and the singlet is lowest.

In the following we will focus on $S=1$ spin chains relevant to the
experiments on Y$_2$BaNiO$_5$.  Numerically the $S=1/2$ chain
end excitations have been observed by various methods : exact
diagonalisation~\cite{Kennedy}, Quantum Monte Carlo (QMC)~\cite{Miyashita},
Density Matrix Renormalization Group (DMRG)~\cite{Polizzi}.
They subsist all through the Haldane phase $-1 < \beta < 1$~\cite{Polizzi}
but as $\beta$ is increased from $0$ to 1 the peak in the onsite
magnetization, $S^z_i$, moves away from the first site of the chain. At
$\beta=1$ a dimerized phase is reached where the $S=1/2$ chain end
excitations cease to exist. At the AKLT point, $\beta=-1/3$, it can be
analytically shown that the decay of the on-site magnetization away from the
end of the chain has a pure exponential form with a length scale exactly equal
to the bulk correlation length.  
Experimentally the $S=1/2$ chain end excitations have been observed in EPR
measurements in $S=1$ chains doped with non magnetic
ions~\cite{hagiwara1,hagiwara2,glarum,EPR,Batista_EPR} and
even in microscopically fractured undoped systems~\cite{Granroth}. 
Initially, experiments measuring the Schottky anomaly in Zn doped
Y$_2$BaNiO$_5$~\cite{ramirez} were interpreted in terms of $S=1$
excitations but later work~\cite{Batista_Cv} showed that an explanation
in terms of $S=1/2$ chain end excitations is also possible when
anisotropy is taken into account. 
These studies have mainly dealt with low temperature (i.e. $T \ll \Delta$)
behavior of these systems where the only relevant
states are the low-lying singlet and triplet.
In the recent experiments by Tedoldi et al~\cite{Tedoldi},
a $^{89}$Y NMR study of doped system
Y$_2$BaNi$_{1-x}$Mg$_{x}$O$_5$ was performed,
clearly indicating the $S=1/2$ chain end excitations for temperatures well
above the gap where many excited states apart from the low-lying
triplet and singlet should be populated.
In order to understand the NMR spectra in this high temperature range
we simulated the isotropic Heisenberg spin 1
chain using a continuous time Quantum Monte
Carlo algorithm at finite temperature and finite field. The QMC method samples the
partition function so that all (excited) states are correctly treated. For the purpose
of understanding the physics at temperatures close to $T\simeq 0$
we have performed Density Matrix Renormalisation Group calculations of the
NMR spectra at $T=0$. Our results for temperatures above the gap are in
very good agreement with the experimental data and clearly show that the 
$S=1/2$ chain end excitations indeed are observable at such high
temperatures. At temperatures below the gap down to $T=0$ we find very well
defined satellite peaks that at $T=0$ only are broadened by the
distribution of chain lengths. 
In the
experiments of Tedoldi~\cite{Tedoldi} et al, no clear satellite
peaks were observed at $T=77K$, well below the gap. This could be due to
a limited resolution at this temperature~\cite{private}. In the experiments of Payen et
al~\cite{Payen_Chi} on doped Y$_2$BaNiO$_5$ the results were interpreted
in terms of the chain end excitations at temperatures down to 4K
and it therefore remains unclear when and if higher dimensional physics
become important in doped Y$_2$BaNiO$_5$.

The paper is organized as follows. In section \ref{method} we discuss
the numerical methods used. Section ~\ref{profiles} presents our results 
for the on site magnetization for different temperatures and fields.
In section \ref{spectra} we turn to a discussion of the NMR spectra
obtained from the magnetization profiles and finally in section
\ref{discussion} we discuss possible explanations why signals from
the $S=1/2$ excitations are not clearly seen below the gap.

\section{Numerical Method}\label{method}
We use a single-cluster continuous time~\cite{Beard} version of the loop
algorithm~\cite{Evertz}. Spins 1 are simulated by two spins $1/2$
with symmetrized boundary conditions~\cite{Troyer,Todo} in the imaginary time
direction. In order to take into account the magnetic field, we use a hybrid
method~\cite{Onishi} where field-dependent loop flip probability and global
flip of worldlines are used. Each Monte Carlo run consists of $10^7$ sweeps
(a measurement is taken every 10 sweeps to avoid autocorrelations) preceded
by $10^5$ sweeps of thermalization. The DMRG~\cite{white2} calculations
were performed keeping $m=243$ states for chain lengths between
L=1,100 and $m=81$ states for chain lengths L=101,501. Only negligible
differences were observed between the results for $m=81$ and $m=243$.
The calculations were performed using the total $z$-component of the
spin, $S^z_{tot}$, and parity, $P$, with respect to a reflection about
the middle of the chain as quantum numbers. All the magnetization
profiles were calculated in the $S^z_{tot}=1, P=-1$ state.

\section{Magnetization profiles}\label{profiles}
\subsection{Influence of temperature}
\begin{figure}
\begin{center}
\epsfig{file=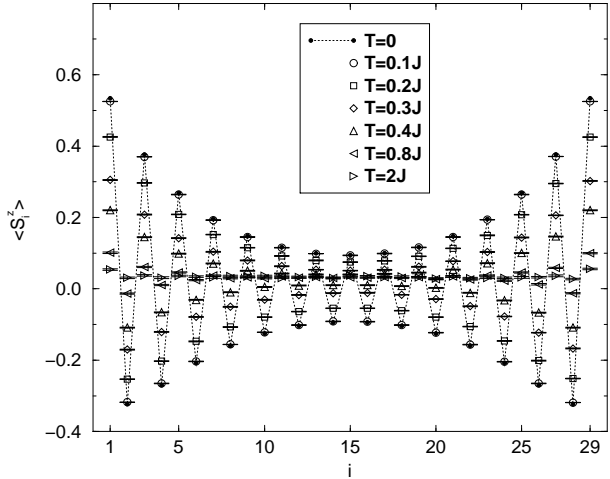,width=8cm}
\caption{Magnetization profiles in the $S^z_{tot}=1$ for an open L=29 chain
for different temperatures. The T=0 profile (filled circles) is obtained by
DMRG, other temperatures by QMC. }
\label{szT}
\end{center}
\end{figure}
The first question we address is the temperature dependence of the
on-site magnetization $S^z_i$. Due to symmetry the on-site magnetization
is always zero in the absence of a magnetic field. In order to compare
to experiments we therefore perform calculations in the sector with
$S^z_{tot}=1$.
In Fig.~\ref{szT} we show the magnetization profile for an open 
chain of 29 spins for various temperatures in the sector
$S^z_{tot}=1$. Filled circles are T=0 results obtained by DMRG, the other
symbols are QMC data. We observe the well-known alternating structure due
to the antiferromagnetic interactions which tends to be diminished as the
temperature is raised.
For the lowest temperature $T=0.1J$ (well below the gap), the profile is
almost identical to the ground-state result. The decay of the on-site
magnetization away from the end of the chain is at $T=0$ known to
be exponential~\cite{SA,Polizzi} with a length scale equal to the bulk
correlation length $\xi\simeq 6$. At high temperature the
staggered magnetization decays rapidly from its value at the end of the
chain corresponding to a shorter correlation length.
At the highest
temperatures, the middle spins (around $i=15$) have become polarized and
show an average magnetization of
$<S_i^z>=\frac{1}{L} \sim 0.034$.  
At temperatures relevant to the experiments of Tedoldi et
al~\cite{Tedoldi}, $T=0.3J-0.8J$, the structure from the 
$S=1/2$ chain end excitations is still clearly visible.
All these results confirm the existence of $S=1/2$ excitations at
temperature above the gap, even if the magnetization at the end
of the chain is significantly reduced above the gap.
\subsection{Influence of magnetic field}
\label{field}
Lifting the constraint of working in the $S^z_{tot}=1$ sector we now
consider the on-site magnetization as a function of the applied magnetic
field at temperatures and fields relevant to the experiments.
We calculated the dependence on the magnetic field of the profiles for two
temperatures : $T=0.1J \, \ll \, \Delta$ and $T=0.526J \, > \, \Delta$.
\begin{figure}
\begin{center}
\epsfig{file=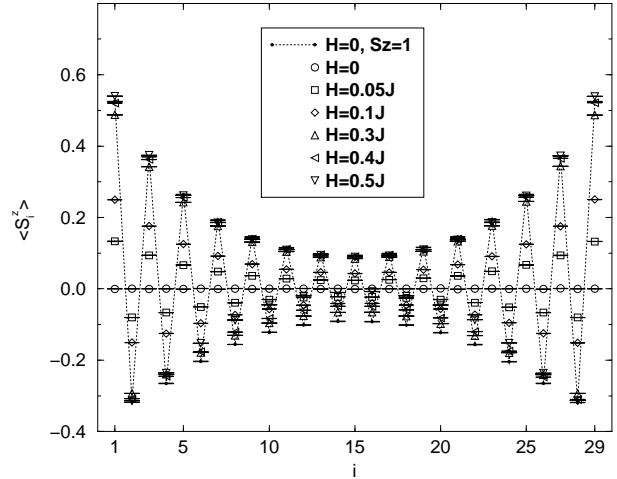,width=8cm}
\caption{Magnetization profiles for an open L=29 chain in different magnetic
fields at $T=0.1J$.}
\label{GammehT01}
\end{center}
\end{figure}

\begin{figure}
\begin{center}
\epsfig{file=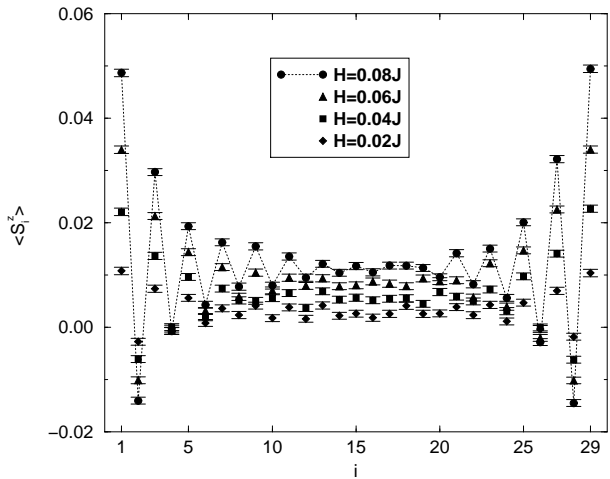,width=8cm}
\caption{Magnetization profiles for an open L=29 chain in different magnetic
fields at $T=0.526J$.}
\label{GammehT2}
\end{center}
\end{figure}

For $T=0.1J$ (figure~\ref{GammehT01}), the zero-field magnetization is zero
for all spins (by symmetry). At this temperature, the only magnetic states
are in the sectors $S_{tot}^z=\pm 1$ of the triplet (which is the lower state
for an odd chain) and are the only ones affected by the field. 
As we increase the field, the $S_{tot}^z=1$ state will be energetically
favored and hence the magnetization starts to take on the alternating structure
showing the presence of the chain end excitations.
At sufficiently large fields we should obtain results comparable to the
results discussed in the previous section obtained in the absence
of a magnetic field but in the $S^z_{tot}=1$ sector.
This is clearly the case and
for $H=0.3J$, we nearly recover the zero-field structure of the
$S_{tot}^z=1$ sector at $T=0.1J$ shown in Fig.~\ref{szT}. 
If the field is too strong (say $H > 0.3J$),
magnetic states from different sectors above the Haldane gap contribute to
the thermal expectation of the on-site magnetization and, even if the
alternating
structure is still present, we don't recover the results from the zero-field
case.  This effect is visible on the $H=0.4J$ data (left triangles). For this field, $<S_i^z>$ for odd spins (which all have a positive expectation
value) is nearly equal to the value in the $S_{tot}^z=1$ sector (filled
circles) but for even spins the value for $H=0.4J$ is greater than in the
$S_{tot}^z=1$ sector (this is particularly clear for the middle even spins).

At higher temperature ($T=0.526J$), slightly above the
gap $\Delta\simeq 0.35-0.4J$ (see figure~\ref{GammehT2}), end effects are
smaller, especially for low fields. At $H=0$, the on-site magnetization is 
zero by symmetry and is not shown in the figure.
Increasing the field, we notice two
effects: the first is simply progressive magnetization of the middle
spins, unaffected by end effects. When we later discuss the NMR spectra
this will result in a positive shift of the NMR spectra for Y atoms close to
the middle of the chain segments, i.e. of the central peak.
The second effect is the progressive 
magnetization of the end spins with increasing field.
The largest value of the field is equal to the field used in the
experiments by Tedoldi et al~\cite{Tedoldi}. Clearly at this field
and temperature there is a signal from the $S=1/2$ chain end
spins.
Since the temperature is large, the correlation length is small
and the end effects are only visible for the first few spins 
away from the chain end.
\subsection{Estimation of correlation length}
As previously mentioned, the on-site magnetization should decay
exponentially away from the end of the chain with a characteristic
length related to the bulk correlation length.
In order to compare to the experiments, we estimated the correlation length for different
temperatures for the longest chains studied in a field of $H=0.08J$.
We fit the exponential decay of the on-site magnetization away from the
end of the chain using the following form:

\begin{eqnarray}
\langle S_i^z\rangle = & & \langle S_1^z\rangle \left((-)^{i-1}e^{-\frac{i-1}{\xi(T)}}+(-)^{L-i}e^{-\frac{L-i}{\xi(T)}}\right) \nonumber\\
 &  & + \langle S_h^z\rangle. 
\end{eqnarray}
Here, $L$ is the length of the chain and the last term $\langle S_h^z\rangle
$ in this
equation corresponds to the magnetization induced by the field at high
temperature. Our results for the correlation length obtained using this
method are shown in table~\ref{Tab} and are in agreement with previous
numerical results~\cite{Todo}~\cite{Kim} and the experimental results
taken from NMR experiments~\cite{Tedoldi}.
\begin{table}
\begin{tabular}{|c|c|c|c|c|}
\hline 
$T(J)$ & 0.993 & 0.702 & 0.526 & 0.27 \\
\hline 
$\xi (T)$ & 1.57 $\pm$ 0.23 & 2.37 $\pm$ 0.37 & 2.97 $\pm$ 0.13 & 4.62 $\pm$
0.11 \\
\hline 
\end{tabular}
\caption{The temperature dependence of the correlation length as
determined from the on-site magnetization profiles in a field of
$H=0.08J$.}\label{Tab}
\end{table}

\section{NMR spectra}\label{spectra}

In the recent $^{89}$Y NMR study by Tedoldi et al~\cite{Tedoldi},
of doped Y$_2$BaNi$_{1-x}$Mg$_{x}$O$_5$ the Y
is sitting roughly at an equal distance between two $S=1$ chains.
The non-magnetic Mg atoms will tend to break the chains into segments
of different lengths and the NMR signal should then be proportional
to the sum of the two nearest Ni spins sitting on different segments
on different chains. Due to the rather large variation in chain
lengths and the resulting variation in the on-site magnetization 
the satellite peaks should be broadened. Averaging over a large
number of chain lengths it is then possible to reconstruct the
observed NMR spectra up to an overall rescaling factor.

In order to perform this calculation we have
simulated open spin 1 segments in a magnetic field of $H=0.08J$,
corresponding to the experimental situation, for 
four temperatures between $\sim 0.2J$ and $\sim 1J$
equal to the experimental values, for chain lengths between L=1,200.
At zero temperature we have computed the profiles in a slightly
different fashion by selecting the $S^z_{tot}=1$ subspace 
for chain lengths between L=1,501 using DMRG techniques.
For very short even length chains, which are quite probable, this
approach might lead to a slight bias since the magnetic field
applied in the experimental situation might not be sufficiently strong
to select the $S^z_{tot}=1$ state as the ground-state. However, we
estimate this effect to be rather small.


Assuming that the distribution of the Mg impurities is uncorrelated
we use a geometric distribution for chain lengths where
the probability of having a chain of length L is given by:
\begin{equation}
p(L)=x(1-x)^L.
\end{equation}
Here $x$ is the impurity concentration. The mean chain length is then 
\begin{equation}
\overline{L}=\sum_{l=0}^{\infty}lx(1-x)^l=\frac{1}{x}-1,
\end{equation}
which, if $x$ is sufficiently small, reduces to
$\overline{L}=\frac{1}{x}$. The non magnetic impurity concentration we used is
$x=5\% $, which corresponds to a mean chain length of 19. 
One should note that the geometric distribution is quite wide and even
though large chain lengths are less probable than short ones they
contribute significantly to the observed NMR spectrum. The most probable
chain length is in fact $L=0$ corresponding to an impurity site and the
possibility of having one or more impurities next to each other is
non-negligible. Hence, in order to obtain reliable results 
we have computed profiles for
chains of long lengths. For $T \neq 0$, we calculated profiles for chains
of lengths 1 up to 200, and for $T=0$ we simulated chains of up to 500
sites~\cite{Proba}.

According to the experimental situation, we randomly choose two lengths of
chains and compute the sum of the two chains magnetizations site by site as
we go along the chains. The two sites whose magnetizations are added at each
step stand for the two next nearest neighbors spins of the NMR probe in the
experiments~\cite{Tedoldi}. When the end of one chain is reached we insert a
site with zero on-site magnetization corresponding to the impurity.
We then select a new chain following the geometrical distribution and
continue to go along the system until the end of another chain is reached
and we repeat the process till the thermodynamic limit is reached.
If the length of the chain is 0 we immediately insert an impurity site
resulting in two or more impurities sitting next to each other.
The NMR spectrum is then computed by plotting the histogram of the
effective field seen on the Y site, i.e. the sum of the on-site
magnetization coming from the two sites sitting closest to the Y on
the two chains. This procedure should be quite close the experimental
situation. In order to compare to the experimental data we 
rescale the y and x axis of the experimental NMR spectrum by a scale factor.
This is done by adjusting
to the first satellite peak (for the x axis we use a scale factor of roughly 
$0.8-0.9$ which is in good agreement with the experimental estimate of
hyperfine interaction). 

\begin{figure}
\begin{center}
\epsfig{file=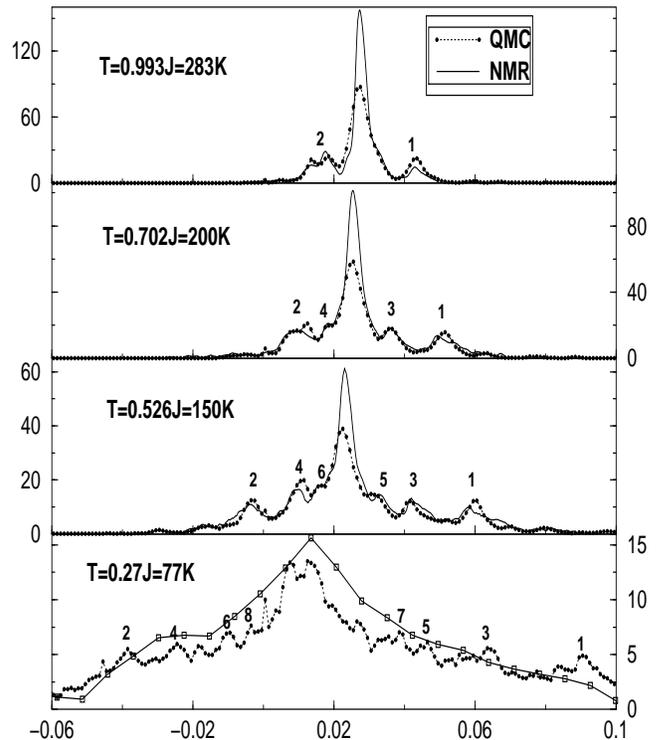,height=9.8cm,width=8.5cm}
\caption{Calculated and experimental NMR spectra for an impurity
concentration of $x=0.05$ at $T=0.993J$, $0.702J$, $0.526J$ and
$0.270J$. The results are obtained using QMC techniques at finite fields and
temperatures for chain lengths of up to $L=200$. Note that the last
experimental spectrum has less resolution than at the other temperatures (the separation between the points corresponds to the experimental resolution).}
\label{fig:spectra}
\end{center}
\end{figure}

\begin{figure}
\begin{center}
\epsfig{file=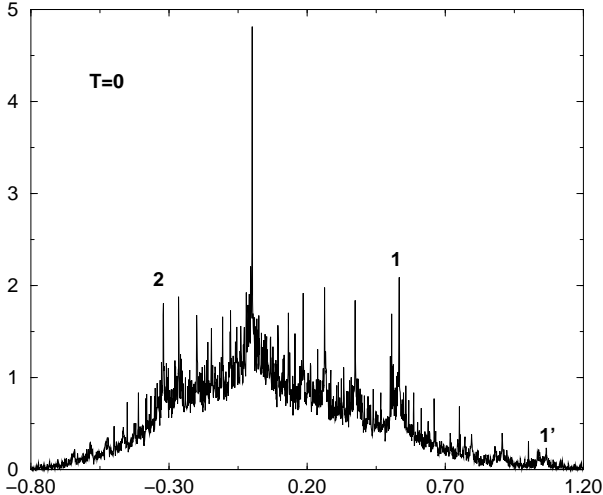,width=8cm}
\caption{Calculated NMR spectra for an impurity concentration of
$x=0.05$ at $T=0$. The results are obtained using DMRG techniques at
zero temperature and zero field in the $S^z_{tot}=1$ subspace.
Chain lengths of up to $L=501$ are used in the calculation.
The peak marked $1'$ corresponds to a site next to {\it two}  chain
ends corresponding to the secondary peaks discussed in the text.}
\label{DMRG}
\end{center}
\end{figure}

In figures~\ref{fig:spectra} and~\ref{DMRG}, we show the results of our 
calculations for the five temperatures and compare them to the experimental 
spectra. The central peak corresponds to the situation where the effective
field at the Y site cancels either due to the fact that the two
contributions are of roughly equal and opposite sign or because the
Y is sitting close to the middle of both of the two chain segments
where the on-site magnetization is very close to zero.
If the Y is closest to the first site of a chain segment it should
experience a positive shift away from the central peak and assuming
that the magnetization on the other chain is close to zero we should
observe the most pronounced satellite peak corresponding to the average
value of the on-site magnetization on the first site. If we have exponentially
localized excitations at the end of the chains the most probably value
of the magnetization on the ``other" chain is close to zero. However,
this is of course not strictly the case, resulting in a broadening of
the peaks. If the Y
is close to the second site of one of the chain we should observe a
satellite peak shifted in the negative direction away from the central peak
but with a smaller absolute shift. It is then possible to identify the various
satellite peaks with the on-site magnetization as one goes away from the
end of the chain. As the distance from the end of the chain is increased
the position of the satellite peak should move closer and closer to
the central peak. Due to the significant broadening of the peaks only
the first few satellite peaks are resolvable. With exponentially
localized excitations at the end of the chains there is also a small
probability that the Y is close to
firsts sites on both chains (for example first-first, first-third ..., or
second-second, second-fourth...). Theoretically, this should result in a
number of secondary peaks beginning at twice the absolute value of the
main peaks, also contributing to a general broadening of the spectrum.
However, the intensity of such peaks should be rather small.
In our calculations at $T=0$, a secondary peak corresponding to the Y
close to the firsts sites on both chains is clearly visible at twice the
value of the first satellite peak (See Fig.~\ref{DMRG}).

The computed spectra are in extremely good agreement with the experimental
results for the three higher temperatures. Note that the central peak is
always greater in the NMR experiments than in our simulations, this might be
due to a different distribution of impurities in the sample than the simple
geometrical distribution (we don't consider correlations between impurities in this model).
There is a central peak for all the temperatures; the position of this peak
is increased as temperature is raised due to the fact that (middle)
spins are more easily magnetized by the magnetic field for high
temperatures, see section~\ref{field}. 
The spectrum is wider at lower temperatures since the magnetization
profiles are much more developed, as discussed in the previous section,
resulting in larger shifts.
For each figure, apart from the central one, several peaks can be observed.
As explained, the shift from the central peak to the others correspond to
the intensity of the magnetization on neighboring sites. 
On the figures, we have labelled the different satellite peaks
according to the proximity to a site on the chain. As can be clearly
seen, the neighboring magnetic structure of a site is more and more well
defined as the temperature is lowered. For example, at the highest temperature
$T=0.993J$, only the two first satellite peaks are resolved, the more
distant satellites cannot be distinguished and contribute to the central
peak. For $T=0.526J$, up to six satellite peaks are resolved, and even more
at the lowest temperature, $T=0.27J$. This temperature is well below the
gap and corresponds to the $T=77K$ experimental results in
Ref.~\onlinecite{Tedoldi}.
In this case the experiment show a single broad
line and the signal from satellite peaks is lost. Tedoldi et
al~\cite{Tedoldi} argues that one possible explanation of this is the large
distribution of the chain lengths which could become more important at
temperatures below the gap. This effect is taken correctly into
account in our calculations and  since we observe well-defined satellite
peaks well below the gap and at $T=0$ the explanation is likely
elsewhere. The resolution of the $T=77K$ spectrum is less than for
the other temperatures~\cite{private} and it is possible that higher
resolution experiments will be capable of resolving the satellite peaks.

Finally, in figure~\ref{DMRG} we show results calculated at $T=0$ using
DMRG techniques. The calculation is performed in zero field in the
$S^z_{tot}=1$ subspace. Many well defined peaks are now visible.
In particular we see secondary peaks corresponding to the case where
the Y is between sites which {\it both} have large on-site
magnetizations ($1'$). 

\section{Discussion}\label{discussion}
Our results clearly indicate that the $S=1/2$ chain excitations should
be visible at temperatures well above the gap in a finite field. 
It is well known that the gap in $S=1$ chains corresponds to a single
magnon branch at $k=\pi$. Around $k=0$ a two-magnon continuum should
begin at energies of $2\Delta$ and a three-magnon continuum at $k=\pi$
should begin at an energy of $3\Delta$~\cite{SA1}. It is natural to
expect that chain end effects are visible until the two-magnon and
three-magnon continuum become significantly populated, i.e. for
temperatures above 300K in agreement with the numerical and
experimental results.

The discrepancy between our theoretical results and the experimental
results for temperatures below the gap could have several explanations.
As suggested by Tedoldi et al~\cite{Tedoldi} anisotropy effects could
be important. However, the single ion anisotropy term $D=8K$ is rather
small and it is not obvious that it would give rise to a large effect.
Such a term would split the low-lying triplet into a doublet and a
singlet and could likely cause a systematic shift in the spectra
rather than a complete smearing of the profile. Still another
explanation could come from the small coupling between the chains.
However, it would seem unlikely that this effect could be important
at temperatures as high as $T=77K$. 
Payen et al~\cite{Payen_Chi} argue that such effects 
become important at temperatures around 4K, much lower than the gap,
$\Delta\simeq 100K$. However, these experiments are concerned with
bulk susceptibility and magnetization which might be a less sensitive
probe.
Another simple explanation is
that below the gap so many secondary peaks appear
that the experimental spectrum appears
as a single broad line. This is not unreasonable considering the results
shown for $T=77K$ in figure~\ref{fig:spectra}.
Lastly, the actual impurity concentration is not known
to a very high precision and a slight modification in this concentration
greatly modifies the observed spectra. High resolution NMR experiments
at low temperatures will be needed to resolve these issues.

\acknowledgements
We would like to acknowledge F. Tedoldi and M. Horvati\'c for many fruitful
discussions and for kindly letting us reproduce the experimental data.


\begin{references}
\bibitem{Proba} According to the geometrical distribution, the probability
to have a chain of length $L>200$ is very small ($< 4.10^{-5}$).
\end{references}
\end{document}